\documentclass[preprint,aps]{revtex4}

\newcommand {\bea}{\begin{eqnarray}}
\newcommand {\eea}{\end{eqnarray}}
\newcommand {\be}{\begin{equation}}
\newcommand {\ee}{\end{equation}}

\usepackage{graphicx}

\begin{document}

\preprint{INT-PUB 04-18}

\title{An Effective Theory for Baryons in the CFL Phase}

\author{A.~Kryjevski$^{1}$ and T.~Sch\"afer$^{2,3}$}

\affiliation{
$^1$ Institute for Nuclear Theory,  University of Washington,
Seattle, WA 98195\\
$^2$Department of Physics, North Carolina State University,
Raleigh, NC 27695\\
$^3$Riken-BNL Research Center, Brookhaven National 
Laboratory, Upton, NY 11973}

\begin{abstract}
We study the effective field theory for fermions in the 
color-flavor locked (CFL) phase of high density QCD. The 
effective theory contains a flavor nonet of baryons interacting 
with a nonet of pseudoscalar Goldstone bosons as well as a 
singlet scalar Goldstone boson. The theory is similar to chiral 
perturbation theory in the one-baryon sector. We explain how 
to incorporate quark mass terms and study the spectrum as a 
function of the strange quark mass. Without meson condensation 
gapless baryon modes appears if the strange quark mass exceeds 
a critical value $m_s^2/(2p_F)=\Delta$, where $p_F$ is the Fermi 
momentum and $\Delta$ is the gap in the chiral limit. We show 
that kaon condensation leads to a rearrangement of the baryon 
spectrum and shifts the critical strange quark mass for the 
appearance of a gapless mode to higher values. 

\end{abstract}
\maketitle

\newpage

\section{Introduction}
\label{sec_intro}

 The color-flavor-locked (CFL) phase is expected to be the 
ground  state of three flavor QCD at very high baryon density
\cite{Alford:1999mk,Schafer:1999fe,Evans:1999at}. In this regime 
the Fermi momentum is much larger than the quark masses and 
flavor symmetry breaking in the QCD lagrangian can be ignored. 
However, at densities that can be achieved in compact stars or 
heavy ion collisions distortions of the pure CFL state due to 
non-zero quark masses cannot be neglected. 

 This problem is usually addressed by studying the effect of a 
non-zero strange quark mass on the solution of a BCS-type gap equation
\cite{Alford:1999pa,Schafer:1999pb,Bedaque:1999nu,Steiner:2002gx,Alford:2002kj,Neumann:2002jm}.
The effect of the strange quark mass on the BCS solution is 
governed by the parameter $m_s^2/(p_F\Delta)$, where $p_F$ is 
the Fermi momentum and $\Delta$ is the gap in the chiral limit. 
It was recently argued that if weak equilibrium and charge 
neutrality are taken into account a new phase of QCD, called
the gapless CFL phase (gCFL), appears at $m_s^2/(2p_F)=\Delta$
\cite{Alford:2003fq,Alford:2004hz,Ruster:2004eg}. This phase is 
characterized by the standard CFL pairing pattern which involves 
all flavors and colors, but the fermion spectrum contains gapless 
excitations. Similar phases were also discussed in the context 
of two flavor QCD  \cite{Shovkovy:2003uu} and cold atomic gases 
\cite{Forbes:2004cr}. The BCS equations predict that for $m_s^2/
(2p_F)=2\Delta$ CFL pairing breaks down completely and a transition 
to a less symmetric phase takes place. 

 In this work we shall study the effect of the strange 
quark mass using an effective field theory of the CFL phase.
The effective field theory for Goldstone modes in the CFL
phase was first discussed in \cite{Casalbuoni:1999wu} and 
mass terms were studied in \cite{Son:1999cm,Schafer:2001za}. 
Bedaque and Sch\"afer showed that if the strange quark mass 
exceeds a critical value $m_s\sim m_u^{1/3}\Delta^{2/3}$ the 
CFL phase undergoes a phase transition to a kaon (or kaon 
and eta) condensed phase 
\cite{Bedaque:2001je,Kaplan:2001qk,Kryjevski:2004cw}. 
A kaon condensate appears because the medium tries to 
reduce its strangeness content as the strange quark mass
is increased. Since the energy gap for strange fermions 
is large but the gap for strange mesons is small, the 
easiest way to reduce strangeness is to form a Bose
condensate of kaons. Kaon condensation is a second order
phase transition. The strangeness and energy density
carried by the kaon condensate are initially small. However,
when $m_s^2/p_F$ approaches the energy gap $\Delta$ the 
strangeness and energy density are of order $\Delta[p_F^2/
(2\pi^2)]$ and $\Delta^2[p_F^2/(2\pi^2)]$, respectively.
These quantities are of the same order as the superfluid
density and the BCS condensation energy. This implies that 
kaon condensation leads to a substantial modification of
the CFL ground state. In the present work we shall investigate
the question how the Bose condensate affects the spectrum 
of fermions. 

\section{QCD at finite isospin density}
\label{sec_mui}

 As a simple illustration for the interaction of a Bose
condensate with fermionic excitations we briefly review 
the computation of the neutrons and proton dispersion 
relations in QCD at finite isospin density \cite{Son:2000xc}. 
The effective lagrangian describing the interaction of low 
energy nucleons with pions is
\be 
\label{leff_mui}
{\cal L} = \bar{\psi}i\gamma^\mu D_\mu \psi 
  - m_{N}\left( \bar{\psi}_L\Sigma \psi_R + h.c.\right)
  + \frac{f_\pi^2}{4} \left\{ {\rm Tr}\left( \nabla_\mu\Sigma
     \nabla_\mu\Sigma^\dagger\right)
    + 2B\left[ {\rm Tr}\left(M\Sigma\right) + h.c. \right]
 \right\} ,
\ee
where $\psi$ is the nucleon field, $\Sigma=\exp(i\pi^a\tau^a/f_\pi)$
is the pion field, and $M$ is the quark mass matrix. Also, 
$m_N$ is the nucleon mass, $f_\pi$ is the pion decay constant
and $B$ determines the pion mass, $B(m_u+m_d)=m_\pi^2$. In the 
presence of a non-zero chemical potential for isospin the 
covariant derivatives are given by $iD_0\psi=i\partial_0\psi-\mu_I
t_3\psi$ and $i\nabla_0\Sigma=i\partial_0\Sigma-\mu_I[t_3,\Sigma]$
with $t_3=\tau_3/2$. A positive isospin chemical potential lowers 
the energy of the neutron and the negative pion and raises the 
energy of the proton and the positive pion, $E_N=m_N \pm |\mu_I|/2$ 
and $E_{\pi^\pm}=m_\pi\pm |\mu_I|$. This suggests that a gapless 
pion excitation appears at $\mu_I= m_\pi$, and a gapless neutron 
appears at $\mu_I\sim 2m_N$.

 The appearance of a gapless pion excitation corresponds 
to the onset of pion condensation. In the pion condensed
phase the expectation value of the chiral field $\Sigma$ 
is given by
\be 
\langle\Sigma\rangle =\cos(\alpha)+i\tau_1\sin(\alpha),
\hspace{1cm}
\cos(\alpha)=\frac{m_\pi^2}{\mu_I^2}.
\ee
The critical chemical potential for a gapless neutron 
to appear in the spectrum is outside the regime of 
validity of the effective theory, $\mu_I\ll \Lambda_\chi
\sim m_\rho$. However, the effective theory can be 
used to study how the pion condensate affects the nucleon
dispersion relation. The nucleon energy in the pion  
condensed phase is $E_N=m_N\pm m_\pi^2/(2|\mu_I|)$. 
Note that this result matches smoothly to the dispersion
relation in the vacuum at $\mu_I=m_\pi$. We observe that 
once pion condensation takes place there is no tendency 
for any of the nucleon states to become gapless. We also 
note that for $\mu_I>m_\pi$ the eigenstates of the Hamiltonian
are mixtures of neutron and proton states. 

 This phenomenon can also be studied using a non-linear
representation for the chiral field. Defining $\Sigma=\xi^2$ 
we can write the nucleon part of the effective lagrangian as
\be
\label{mui_nlin}
 {\cal L} = \bar{N} \gamma^\mu\left(iD_\mu
 -g_A \gamma_5{\cal A}_\mu \right) N - m_N\bar{N}N,
\ee
where $N_R=\xi\psi_R$, $N_L=\xi^\dagger\psi_L$. The 
covariant derivative is $iD_\mu N = (i\partial_\mu 
- {\cal V}_\mu) N$ and the vector and axial-vector currents 
are given by
\bea
{\cal V}_\mu &=& -\frac{i}{2} \left\{ 
  \xi \left(\partial_\mu
       -i\delta_{\mu,0}\mu_I\frac{\tau_3}{2}\right)\xi^\dagger + 
  \xi^\dagger \left(\partial_\mu
       -i\delta_{\mu,0}\mu_I\frac{\tau_3}{2}\right)\xi  \right\}, \\
{\cal A}_\mu &=& -\frac{i}{2} \left\{ 
  \xi \left(\partial_\mu
       -i\delta_{\mu,0}\mu_I\frac{\tau_3}{2}\right)\xi^\dagger - 
  \xi^\dagger \left(\partial_\mu
       -i\delta_{\mu,0}\mu_I\frac{\tau_3}{2}\right)\xi 
  \right\}.
\eea
The linear chiral theory given in equ.~(\ref{leff_mui}) leads 
to $g_A=1$, but chiral symmetry does not determine the value 
of the axial coupling constant. The covariant derivative 
takes into account the interaction of the pionic iso-vector 
current with the nucleon. In the pion condensed phase there 
is a repulsive interaction between the iso-vector charge of 
the neutron and the pion condensate which cancels the shift 
of the neutron mass due to the chemical potential. In addition 
to the vector interaction, which is diagonal in flavor, there 
is off-diagonal mixing caused by the axial-vector interaction. 
This interaction, however, is suppressed by the large mass of 
the nucleon and does not affect the dispersion relation at 
leading order in the chiral expansion.

\section{Effective field theory for baryons in the CFL phase}
\label{sec_cfl}

 In the CFL phase we start from the high density effective 
lagrangian \cite{Hong:2000ru,Nardulli:2002ma,Schafer:2003jn}
\bea
\label{l_cfl1}
 {\cal L} &=& {\rm Tr}\left(\psi^\dagger_L (iv\cdot D) \psi_L
                           +\psi^\dagger_R (iv\cdot D) \psi_R\right) 
   -{\rm Tr}\left(\psi^\dagger_L \frac{MM^\dagger}{2p_F}\psi_L
                 +\psi^\dagger_R \frac{M^\dagger M}{2p_F}\psi_R \right) 
 \nonumber \\
 & &  \mbox{} + \frac{\Delta}{2} \left\{ {\rm Tr} \left( X^\dagger \psi_L
                       X^\dagger \psi_L \right) 
  - \left[ {\rm Tr}\left( X^\dagger\psi_L\right)\right]^2 
    + h.c. \right\} \nonumber \\
 & &  \mbox{} - \frac{\Delta}{2} \left\{ {\rm Tr}\left( Y^\dagger \psi_R
                       Y^\dagger \psi_R\right) 
  - \left[ {\rm Tr}\left(Y^\dagger\psi_R\right)\right]^2 
    + h.c. \right\}.
\eea
Here, $\psi_{L,R}$ are left and right-handed quark fields
which transform as
\be 
 \psi_L\to L\psi_L C^T, \hspace{1cm}
 \psi_R\to R\psi_R C^T,
\ee
under chiral transformations $(L,R)\in SU(3)_L\times SU(3)_R$ 
and color transformations $C\in SU(3)_C$. Also, $D_\mu$ is 
the $SU(3)_C$ covariant derivative, $v_\mu=(1,\vec{v})$ is the
local Fermi velocity. We have suppressed the spinor indices
and defined $\psi\psi=\psi^\alpha\psi^\beta {\cal C}^{\alpha\beta}$, 
where ${\cal C}$ is the charge conjugation matrix. The traces run over 
color or flavor indices. $M$ is the quark mass term, and $X,Y$ 
are fields that transform as
\be
 X\to L X C^T, \hspace{1cm}
 Y\to R Y C^T.
\ee
We will assume that the vacuum expectation value is $\langle X 
\rangle =\langle Y \rangle = 1$. This corresponds to the CFL gap 
term $\Delta(\psi_L)^a_i(\psi_L)^b_j(\delta^a_i\delta^b_j-\delta^a_j
\delta^b_i)-(L\leftrightarrow R)$. For simplicity we have assumed
that the gap term is completely anti-symmetric in flavor. This 
assumption is satisfied to leading order in the strong coupling 
constant. Our results are easily generalized to allow for a 
small admixture of a flavor symmetric gap. We will derive the 
effective lagrangian in the chiral limit $M=M^\dagger=0$. The 
effect of the leading order mass term in equ.~(\ref{l_cfl1}) 
is easily recovered using the symmetries of the effective theory. 

We can redefine the fermion fields according to 
\be 
\chi_L \equiv \psi_L X^\dagger , \hspace{1cm}
\chi_R \equiv \psi_R Y^\dagger.
\ee
In terms of the new fields the lagrangian takes the form 
\bea
{\cal L} &=& {\rm Tr}\left( \chi^\dagger (iv\cdot\partial ) \chi\right)
 \nonumber \\ 
 & & \hspace{0.5cm}\mbox{}
  -i\,  {\rm Tr}\left( \chi_L^\dagger \chi_L
   X v^\mu \left( \partial_\mu -iA_\mu^T \right)X^\dagger \right) 
  -i\,  {\rm Tr}\left( \chi_R^\dagger \chi_R
   Y v^\mu \left( \partial_\mu -iA_\mu^T \right)Y^\dagger \right) 
   \nonumber \\ 
 & &  \hspace{0.5cm}\mbox{} + \frac{\Delta}{2} \left\{ 
     {\rm Tr} \left( \left(\chi_L\chi_L \right) 
   - \left[ {\rm Tr}\left(\chi_L\right)\right]^2 \right)  
   - \left( {\rm Tr} \left(\chi_R\chi_R \right) 
   - \left[ {\rm Tr}\left(\chi_R\right)\right]^2 \right) 
     + h. c.   \right\} . 
\eea
We can now integrate out fermion loops. Our aim is to derive 
an effective theory for Goldstone bosons interacting with heavy 
fermions. From the construction of the theory it is clear that 
the heavy fermions can no longer occur in loops. Casalbuoni and
Gatto showed that fermion loops generate a kinetic term for
the chiral fields $X$ and $Y$ \cite{Casalbuoni:1999wu}
\be
\label{l_higgs}
{\cal L}=-\frac{f_\pi^2}{2}\left\{
  {\rm Tr}\left( (X^\dagger D_0 X)^2-v_\pi^2(X^\dagger D_i X)^2\right)
+ {\rm Tr}\left( (Y^\dagger D_0 Y)^2-v_\pi^2(Y^\dagger D_i Y)^2\right)
\right\}.
\ee
Here we have ignored the flavor singlet components of $X$ and $Y$ 
since they are not important for the baryon spectrum.  The low 
energy constants $f_\pi$ and $v_\pi$ were calculated in 
\cite{Son:1999cm}, see also 
\cite{Bedaque:2001je,Beane:2000ms,Zarembo:2000pj,Rho:1999xf}.
The results are
\be 
\label{f_pi}
f_\pi^2=\frac{21-8\log(2)}{18}
  \left(\frac{p_F^2}{2\pi^2}\right), \hspace{1cm}
v_\pi^2= \frac{1}{3}.
\ee
The gluon field acquires a mass due to the Higgs mechanism. 
This implies that the gluon is heavy and can be integrated
out. Using equ.~(\ref{l_higgs}) we get
\be 
 A_\mu^T = \frac{i}{2}\left( X^\dagger \partial_\mu X 
  + Y^\dagger \partial_\mu Y \right) + \ldots
\ee
This result can be substituted back into the lagrangian for 
the heavy degrees of freedom. The result can be simplified 
by using the gauge choice $X=Y^\dagger=\xi$. The Goldstone 
boson field is given by $\Sigma=XY^\dagger=\xi^2$. The field
$\xi$ transforms as $\xi\to L\xi U^\dagger =U \xi R^\dagger$ 
where $U(x)$ is defined by the transformation law $\Sigma\to L
\Sigma R^\dagger$ together with $\Sigma=\xi^2$. For global 
flavor transformations $L=R=V$ we have $U(x)=V$. In general, 
$U(x)$ depends on the Goldstone boson field. We will also 
perform a further field redefinition for the fermion field, 
\be 
 N_L=\xi^\dagger \chi_L\xi, \hspace{1cm}
 N_R=\xi \chi_R \xi^\dagger 
\ee
The fermion field transforms as $N\to U N U^\dagger$. In 
agreement with the general analysis of the symmetries
of the CFL phase we find that the fermion excitations 
are described by a field that transforms as a nonet of 
flavor $SU(3)$. The field $N$ transforms under $U(1)_B$ 
as $N\to \exp(i\phi)N$ whereas the original quark field 
transformed as $\psi\to\exp(i\phi/3)\psi$. We conclude 
that $N$ describes a baryon nonet. The effective lagrangian 
for the baryons is 
\bea 
\label{l_bchpth1}
{\cal L} &=& {\rm Tr}\left(N^\dagger iv^\mu D_\mu N\right) 
- D{\rm Tr} \left(N^\dagger v^\mu\gamma_5 
               \left\{ {\cal A}_\mu,N\right\}\right)
- F{\rm Tr} \left(N^\dagger v^\mu\gamma_5 
               \left[ {\cal A}_\mu,N\right]\right)
 \nonumber \\
 & &  \hspace{0.5cm}\mbox{} + \frac{\Delta}{2} \left\{ 
     \left( {\rm Tr}\left(N_LN_L \right) 
   - \left[ {\rm Tr}\left(N_L\right)\right]^2 \right)  
   - \left( {\rm Tr} \left(N_RN_R \right) 
   - \left[ {\rm Tr}\left(N_R\right)\right]^2 \right) 
     + h. c.  \right\},
\eea
with $D_\mu N=\partial_\mu N +i[{\cal V}_\mu,N]$ and, to 
leading order in $\alpha_s$ and $1/p_F$, $D=F=1/2$. The vector 
and axial-vector currents are given by 
\be
 {\cal V}_\mu = -\frac{i}{2}\left\{ 
  \xi \partial_\mu\xi^\dagger +  \xi^\dagger \partial_\mu \xi 
  \right\}, \hspace{1cm}
{\cal A}_\mu = -\frac{i}{2} \xi\left(\partial_\mu 
    \Sigma^\dagger\right) \xi . 
\ee
Up to the presence of the gap terms, and some kinematic 
differences, this result has the same structure as the 
lagrangian of flavor $SU(3)$ baryon chiral perturbation 
theory. This is in agreement with the general arguments 
presented in \cite{Schafer:1999ef} that the CFL phase is 
indistinguishable from a hyper-nuclear phase. We note, 
in particular, that the interaction of the baryons
with the vector current is entirely dictated by chiral
symmetry. 

 Symmetry arguments can also be used to determine the 
leading mass terms in the effective lagrangian. Bedaque
and Sch\"afer observed that $X_L=MM^\dagger/(2p_F)$ and
$X_R=M^\dagger M/(2p_F)$ enter equ.~(\ref{l_cfl1}) like
left and right handed flavor gauge fields. We can make 
the baryon effective lagrangian invariant under this 
gauge symmetry by introducing the covariant derivatives 
\bea
\label{V_X}
 D_0N &=& \partial_0 N+i[\Gamma_0,N], \hspace{0.5cm}
 \Gamma_0 = -\frac{i}{2}\left\{ 
  \xi \left(\partial_0+ iX_R\right)\xi^\dagger + 
  \xi^\dagger \left(\partial_0+iX_L\right) \xi 
  \right\}, \\ 
\nabla_0\Sigma &=& \partial_0\Sigma+iX_L\Sigma-i\Sigma X_R
\eea
The terms involving $X_L,\ X_R$ are corrections of order
$M^2/(p_F\Delta)$. The CFL phase has an approximate $U(1)_A$
symmetry which forbids terms linear in $M$. We performed a 
detailed study of mass corrections in high density QCD in 
\cite{Schafer:2001za}. We showed, in particular, that in 
addition to the $X_{L,R}$ terms included in equ.~(\ref{l_cfl1}) 
there are other $O(M^2)$ terms which generate corrections of 
$O(M^2/p_F^2)$. These terms are suppressed by a factor of 
$\Delta/p_F$ compared to the terms in equ.~(\ref{V_X}). 

 There are two types of $O(M^2/p_F^2)$ terms. The first
category contains $N^\dagger N$ terms that are of the same
structure as higher order corrections to the baryon sigma
term in ordinary baryon chiral perturbation theory. An 
example is the term ${\rm Tr}(N^\dagger \phi_+^2N)$, 
where $\phi_\pm=\phi_L\pm \phi_R$ and $\phi_L=\xi M^\dagger 
\psi, \phi_R=\xi^\dagger M\xi$.   The second category 
consists of $O(M^2)$ corrections to the gap. Chiral 
symmetry allows both terms like ${\rm Tr}(\phi_+^2)
{\rm Tr}(NN)$ that do not violate the flavor symmetry 
of the gap term, and terms like ${\rm Tr}(\phi_+N\phi_+ N)$ 
which break the symmetry. We note, however, that if we
restrict ourselves to flavor anti-symmetric gap terms 
then we find that in the trivial vacuum $\xi=1$ there 
is no flavor symmetry breaking in the gaps.

\section{Baryon spectrum}
\label{sec_spec}

 In the following we shall restrict ourselves to terms of 
order $O(M^2/(p_F\Delta))$. The baryon spectrum in the CFL 
phase is determined by 
\be 
{\cal L} = {\rm Tr}\left(N^\dagger iv^\mu D_\mu N\right)
 + {\rm Tr}\left(N^\dagger \gamma_5 \rho_A N\right)
 +\frac{\Delta}{2} \left\{ {\rm Tr}\left(N\gamma_5 N\right) -
  {\rm Tr}\left(N\right)\gamma_5{\rm Tr}\left(N\right)+ h.c.\right\},
\ee
where we have used $D=F=1/2$. The covariant derivative is 
$D_0N=\partial_0N+i[\rho_V,N]$  with 
\be 
\rho_{V,A} = \frac{1}{2}\left\{ 
  \xi \frac{M^\dagger M}{2p_F}\xi^\dagger \pm 
  \xi^\dagger \frac{MM^\dagger}{2p_F} \xi 
  \right\}.
\ee
We shall study the energy of particle and hole 
excitations as a function of the effective chemical
potential $\mu_s=m_s^2/(2p_F)$, neglecting terms 
of order $m_u^2,m_d^2$. We will focus on the minimum
energy required to create an excitation. 

 In the CFL vacuum we have $\xi=1$. As a function of 
$\mu_s$ the excitation energy of proton and neutron particles, 
as well as $\Xi^-,\Xi^0$ holes, is lowered, $\omega_{p,n}=
\Delta-\mu_s$, while the energy of $\Xi^-,\Xi^0$ particles 
and proton and neutron holes is raised, $\omega_{\Xi}=\Delta
+\mu_s$. All other excitation energies are independent of
$\mu_s$. As a consequence we find gapless $(p,n)$ and 
$(\Xi^-,\Xi^0)^{-1}$ excitations at $\mu_s=\Delta$. This 
result is in agreement with the fermion spectrum in 
``microscopic'' calculations \footnote{The equivalence 
between our results and the results of \cite{Alford:2004hz}
can be checked using the following dictionary between 
the quark and hadron description: $(rd,gu,rs,bu,gs,bd)=
(\Sigma^-,\Sigma^+,\Xi^-,p,\Xi^0,n)$. The remaining 
$(ru,gd,bs)$ states are linear combinations of the 
neutral baryons $(\Sigma^0,\Lambda^8,\Lambda^0)$.}. Note 
that in microscopic calculations the energy shift is 
a combination of the effective chemical potential $\mu_s$
and a color chemical potential required to keep the system
color neutral. The effective theory is formulated in terms
of gauge invariant variables and no color chemical 
potential is required. The physical effect of the color
chemical potential is taken into account through the fact 
that, after integrating out the gauge field, the effective 
chemical potential acts on fermion fields in the adjoint 
rather than the fundamental representation of flavor $SU(3)$. 

 When the effective chemical potential $\mu_s$ exceeds the 
mass of the kaon the ground state changes and a Bose condensate
of kaons is formed. If isospin was an exact symmetry the 
energy of a $K^+$ or $K^0$ condensate would be exactly 
degenerate. In the real world, because $m_u<m_d$, and because
of the constraint of electric charge neutrality, a condensate
of neutral kaons is favored. In the $K^0$ condensed phase we have 
\begin{figure}
\includegraphics[width=14cm]{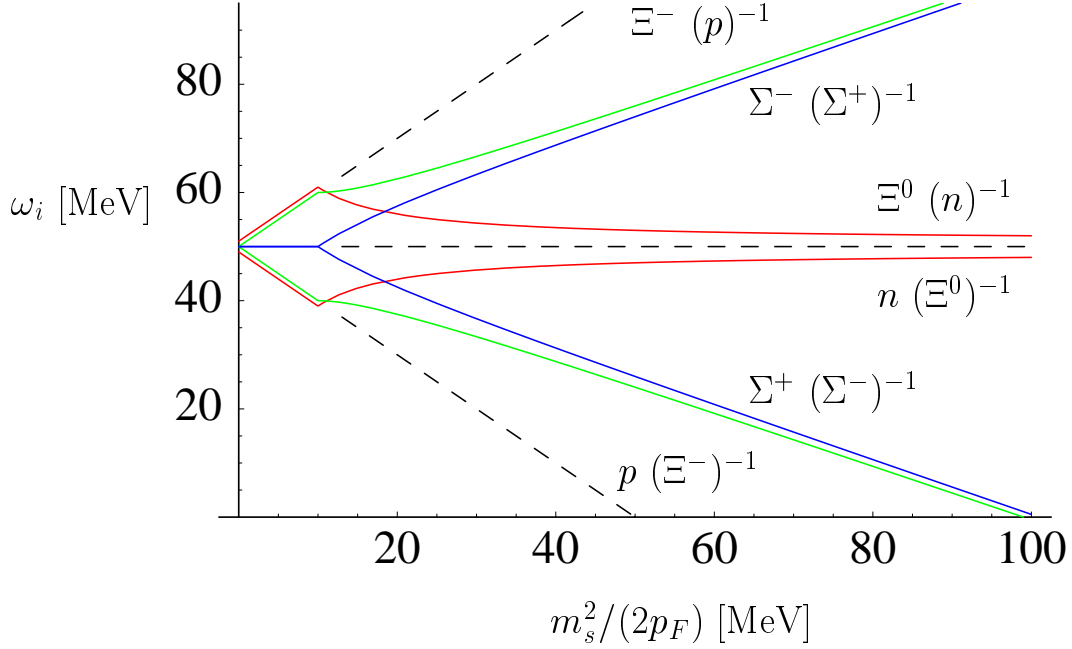}
\caption{Baryon spectrum in the CFL phase as a function of 
$m_s^2/(2p_F)$ for $F=D=0$. The figure shows the minimum 
energy required to create a particle or hole excitation. We
have chosen the gap in the chiral limit to be $\Delta=50$ MeV
and the critical value of $\mu_s$ for $K^0$ condensation to be 
10 MeV. The dashed curves show the spectrum without $K^0$ 
condensation. In that case, gapless $(p,n)$ particle and $(\Xi^-,\Xi^0)$ 
hole excitations appear at $m_s^2/(2p_F)=\Delta$. The full lines
show the spectrum with kaon condensation included. In the $K^0$ 
condensed phase gapless $(p,\Sigma^+)$ particle and $(\Xi^-,\Sigma^-)$
hole excitations appear at $m_s^2/(2p_F)=2\Delta$.}
\label{fig_spec}
\end{figure}
\be 
\xi_{K^0} = \left( \begin{array}{ccc}
 1 &      0          &     0            \\
 0 & \cos(\alpha/2)  & i \sin(\alpha/2) \\
 0 & i\sin(\alpha/2) & \cos(\alpha/2) 
\end{array} \right),
\ee
with $\cos(\alpha)=m_K^2/\mu_s^2$. In the weak coupling limit we 
have $m_K^2 = 3m_u(m_d+m_s)\Delta^2/(\pi^2 f_\pi^2)$ where $f_\pi$ is 
given in equ.~(\ref{f_pi}). The vector charge density $\rho_V$ remains
diagonal in the $K^0$ condensed phase, while the axial charge density
$\rho_A$ causes flavor mixing. We first analyze the spectrum in 
the absence of flavor mixing, i.e. for $F=D=0$. The isospin and 
hypercharge components of the vector potential $\rho_V$ are given 
by $\rho_{I_3}=-\mu_s(1/2-\cos(\alpha)/2)$ and $\rho_{Y}=-\mu_s
(1/4+3\cos(\alpha)/4)$. The excitation energies can now be determined 
from the $(I_3,Y)$ quantum numbers of the baryon nonet. We find
\begin{figure}
\includegraphics[width=14cm]{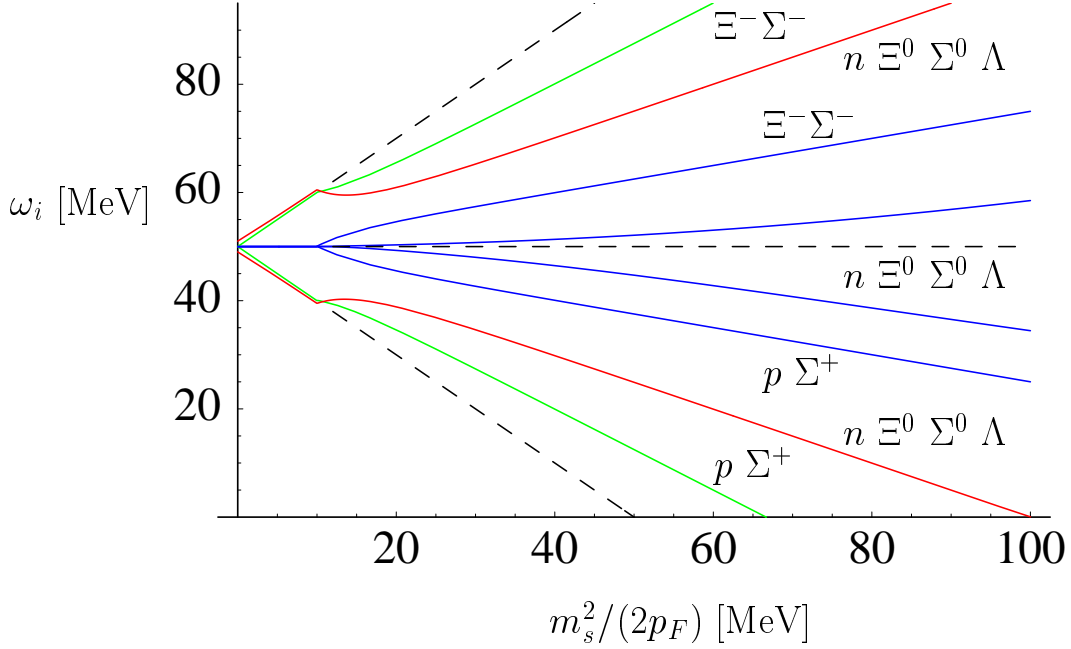}
\caption{Baryon spectrum in the CFL phase as a function of 
the parameter $m_s^2/(2p_F)$ for $F=D=1/2$. The parameters
were chosen as in Fig.~\ref{fig_spec_2}.}
\label{fig_spec_2}
\end{figure}
\bea
 \omega_{n\ ,\Xi^0} &=& \Delta \mp  \mu_s\cos(\alpha), \nonumber \\
 \omega_{\Sigma^+,\Sigma^-} &=& \Delta \mp 
            \frac{\mu_s}{2}(1-\cos(\alpha)), \\
 \omega_{p\ ,\Xi^-} &=& \Delta \mp 
            \frac{\mu_s}{2}(1+\cos(\alpha)). \nonumber 
\eea
The results are shown in Fig.~\ref{fig_spec}. We observe that 
the $(n,\Xi^0)$ system is exactly analogous to the $(n,p)$ system 
in the case of QCD at finite isospin. After $K^0$ condensation 
takes place there is no tendency for neutron particle or $\Xi^0$
holes to become gapless. However, even with $K^0$ condensation
included, the proton excitation energy still drops as a function
of $\mu_s$. In addition to that a new light mode, the $\Sigma^+$,
appears. The reason is that the $K^0$ condensate reduces the 
number of strange quarks at the expense of introducing extra 
down quarks into the system. As a consequence, there is an 
attractive interaction for baryons that have an excess of up
quarks over down quarks. We note that nevertheless the 
critical effective chemical potential for which the proton 
and $\Sigma^+$ become gapless is twice the corresponding value 
in the CFL vacuum. 

 When the axial coupling is taken into account there is mixing 
in the $(p,\Sigma^+,\Sigma^-,\Xi^-)$ and $(n,\Sigma^0,\Xi^0,\Lambda^8,
\Lambda^0)$ sector. As a consequence the spectrum is more complicated.
For $\mu_s\gg m_K$ the spectrum of the charged modes is given by
\be
\omega_{p\Sigma^\pm\Xi^-}= \left\{
 \begin{array}{l}
 \Delta \pm \frac{3}{4}\mu_s, \\
 \Delta \pm \frac{1}{4}\mu_s,
\end{array}\right.  \hspace{1.55cm}
(p_m)_{p\Sigma^\pm\Xi^-}=\left\{
 \begin{array}{l}
  p_F + \frac{1}{4}\mu_s, \\
  p_F - \frac{1}{4}\mu_s,
\end{array}\right. 
\ee
where $p_m$ is the momentum at which the dispersion 
relation has a minimum. The spectrum of the neutral modes is
\be
\omega_{n\Sigma^0\Xi^0\Lambda} = \left\{
 \begin{array}{l}
   \Delta \pm \frac{1}{2}\mu_s +O(\mu_s^2),\\ 
   \Delta  + O(\mu_s^2), \\
   2\Delta + O(\mu_s^2).
 \end{array} \right.
 \hspace{1cm}
 (p_m)_{n\Sigma^0\Xi^0\Lambda}=p_F.
\ee
Numerical results for the eigenvalues are shown in Fig.~\ref{fig_spec_2}. 
We observe that mixing within the charged and neutral baryon sectors leads 
to level repulsion. There are two modes that become light in the CFL window 
$\mu_s\leq 2\Delta$. One mode is a linear combination of proton and 
$\Sigma^+$ particles, as well as $\Xi^-$ and $\Sigma^-$ holes, and 
the other mode is a linear combination of the neutral baryons $(n,
\Sigma^0,\Xi^0,\Lambda^8,\Lambda^0)$.

. 

\section{Conclusions}
\label{sec_sum}

 We have constructed an effective field theory for 
baryons in the CFL phase. We have shown that to 
$O(M^2/(p_F\Delta))$ mass terms in the effective 
theory are completely determined by the symmetries
of high density QCD. We have computed the baryon 
spectrum as a function of the strange quark mass
and showed that kaon condensation substantially 
modifies the baryon spectrum. Both with and without 
kaon condensation we find that light baryon modes 
appear in the spectrum when $m_s^2\sim p_F\Delta$. 
We note, however, that kaon condensation increases
the estimate for the critical strange quark mass
at which gapless modes appear. We also emphasize
that a reliable study of the possibility of a 
gapless phase requires a resummation of terms of 
all orders in $(M^2/(p_F\Delta))$. Progress in this 
direction is reported in \cite{Andrei}.

Acknowledgments: We wish to thank Mark Alford, David Kaplan
and Sanjay Reddy for helpful discussions. We thank Krishna
Rajagopal for pointing out a discrepancy between an earlier
draft and reference \cite{Andrei} and Mei Huang and Igor 
Shovkovy for communications regarding their work on the 
stability of the g2SC phase \cite{Huang:2004bg}. A different 
perspective on baryons in the CFL phase can be found 
in \cite{Hong:1999dk}. Effective theories similar to the one 
considered in this paper were constructed for QCD at zero 
density \cite{Wetterich:2000pp} and for the 2SC phase 
\cite{Casalbuoni:2000cn}. The work is supported in part by 
the US Department of Energy grant DE-FG02-00ER41132 (A.K.) and 
DE-FG-88ER40388 (T.S.).


\end{document}